\documentclass{article}

\usepackage{graphicx} 
\usepackage{rotating}
\usepackage{array}
\usepackage{amsmath}
\usepackage{amssymb}
\usepackage{amsthm}
\usepackage{url}
\usepackage{breakurl}
\usepackage{multirow}
\usepackage{chemformula}
\usepackage{indentfirst}
\usepackage{mdwlist}
\usepackage{setspace}
\usepackage{mathtools}
\usepackage{float}

\usepackage[T1]{fontenc}

\title{Fast exact computation of the $k$ most abundant isotope peaks with layer-ordered heaps}

\begin{document}

\author{Patrick Kreitzberg\\
  University of Montana\\
  Department of Computer Science
  \and
  Jake Pennington\\
  University of Montana\\
  Department of Computer Science
  \and
  Kyle Lucke\\
  University of Montana\\
  Department of Computer Science\\
  \and
  Oliver Serang\\
  University of Montana\\
  Department of Computer Science\\
  Oliver.Serang@umontana.edu
}

\maketitle

\begin{abstract}
\noindent The theoretical computation of isotopic distribution of compounds is
  crucial in many important applications of mass spectrometry,
  especially as machine precision grows. A considerable amount of good
  tools have been created in the last decade for doing so. In this
  paper we present a novel algorithm for calculating the top $k$ peaks
  of a given compound. The algorithm takes advantage of layer-ordered
  heaps used in an optimal method of selection on $X+Y$ and is able to
  efficiently calculate the top $k$ peaks on very large molecules.
  Among its peers, this algorithm shows a significant speedup on
  molecules whose elements have many isotopes. The algorithm obtains a
  speedup of more than 31x when compared to $\textsc{IsoSpec}$ on
  \ch{Au2Ca10Ga10Pd76} when computing 47409787 peaks, which covers
  0.999 of the total abundance.
\end{abstract}

\section{Introduction}
Calculating the theoretical isotopic distribution of compounds is a
valuable tool in mass spectrometry (MS), but is also a difficult
combinatorics problem to solve without doing some sort of statistical
approximation. Theoretical isotopic distribution may be used in
targeted screening\cite{ruff:quantitative, singer:rapid}, identifying
unknown metabolites\cite{ji:deep}, and is useful in general MS
workflow \cite{sturm:openms} among others.

There have been multiple methods developed in the past decade focused
on more efficient methods of calculating the isotope distribution
\cite{ipsen:efficient,loos:accelerated,lacki:isospec,wang:efficient,sadygov:poisson}.
Some, such as \textsc{IsoSpec}, take a combinatorial approach to the
problem and are able to solve for the exact mass and abundance, others
such as Sadygov's method uses a more statistical approach by using an
approximation of binomial and multinomial distributions. Despite the
results being the same, many of these methods taken different input
parameters. \textsc{IsoSpec}, among others, asks the user to provide
some minimum threshold $p$ so that the returned list of peaks has a
total abundance of at least $p$, other methods simply generate all
peaks above a certain abundance threshold. In 2019, Wang \emph{et al.}
compared four of the top algorithms: \textsc{IsoSpec},
\textsc{enviPat}, \textsc{ecipex} and their own \textsc{isoVector}.
They found \textsc{IsoSpec} to consistently be the fastest of the
four.

\textsc{IsoSpec} works by first calculating the most abundant
subisotopologues (all instances of the same element in a compound, for
example \ch{H2} and \ch{O} are two subisotopologues of water) and then
combining the subisotopologues to form whole isotopologues. The
isotopologues are then put into a FIFO queue and when one is popped,
if it exceeds some threshold, it will be appended to the output and
its neighbors (isotopologues who differ by one incrementing one
isotope of an element and decrementing another isotope of the same
element by one) are inserted into the queue, else the isotopologue
will be considered for the next layer. Each threshold for the FIFO
queue creates a new layer of isotopologues, the cumulative output of
all layers is $\in O(\cdot)$ of the optimal output; however, in
practice \textsc{IsoSpec} may produce significantly more isotopologues
than are necessary and so the output must be pruned before being
reported.

Once two subisotopologues have been calculated, selecting the top $k$
isotopologues of the resulting compound created by merging the two
subisotopologues is the same as selecting the top $k$ terms in a
Cartesian product of two lists. Since the Cartesian product uses
addition, but the abundances should be multiplied, the log of the
abundances are used in the $X+Y$ selection. There are multiple methods
for selecting the top $k$ terms in the Cartesian product on $X+Y$ in
optimal time\cite{kaplan:selection,serang:optimal}. Serang's optimal
method utilizes layer-ordered heaps (LOH), a data structure which is
contiguous in memory and partitioned into layers. The size of each
layer grows exponentially according to some growth rate $\alpha$ and
each value in layer $i$ is less than or equal to each value in layer
$i+1$. The advantage of a LOH is that an array may be LOHified in
$O(n)$ time where sorting the array is $\in \Omega(n\log(n))$.

In this paper, we present a method of efficiently calculating the top
$k$ isotopologues of a compound using LOHs. The method does not
approximate, bin or round any numbers. In fact, perfect numbers could
be used if machine precision allowed. This is achieved by building a
balanced binary tree out of two separate types of nodes. The leaf
nodes calculate the most abundant subisotopologues by performing
selection on a multinomial, and outputing the results in a
layer-ordered format. Other nodes solve pairwise selection problems,
using a modified version of Serang's method, which are the outputs of
either the subisotopologues or pairs of already merged
subisotopologues. The outputs of the root of the tree are the most
abundant isotopologues of the given compound. The output will not be
sorted but will instead be in layer-ordered. This method has been
packaged as a freely available \texttt{C++} software package entitled
\textsc{NeutronStar}.

\section{Methods}
Here, we present several methods which, given an input of $k$ and a
chemical formula, may be combined to calculate the top $k$ most
abundant isotopologues of the given chemical.

\subsection{Selection on a multinomial}
Calculating the subisotopologues of a compound is equivalent to
calculating a polynomial
$(b_1\cdot X_1^{\beta_1} + b_2\cdot X_1^{\beta_2} + \cdots)^{q_1}$
where $b$ is the log-abundance and $\beta$ the mass of an isotope. For
an element with four Carbon, the corresponding polynomial is
$(\ch{^{12}C}\cdot X^{12.0} + \ch{^{13}C}\cdot X^{13.003})^4$. Then,
combining two subisotopologues is the equivalent of multiplying two
polynomials (\emph{e.g.}, \ch{H3C4} is
$(\ch{^{12}C}\cdot X^{12.0} + \ch{^{13}C}\cdot X^{13.003})^4 \cdot
(\ch{^{1}H}\cdot X^{1.008} + \ch{^{2}H})^3$.

When calculating a polynomial such as $(c_1\cdot X + c_2\cdot Y)^2$
there will be terms which can be merged together:
$(c_1\cdot X + c_2\cdot Y)^2 = (c_1^2\cdot X^2 + c_1c_2\cdot X\cdot Y
+ c_2c_1\cdot Y\cdot X + c_2^2\cdot Y^2) = (c_1^2\cdot X^2 +
2c_1c_2\cdot X\cdot Y c_2^2\cdot Y^2)$. For larger polynomials (both
in the power and number of terms) there will be many terms which may
combine with each other. A significant speed-up is to be found if only
one of these terms is calculated then multiplied by the appropriate
multinomial coefficient (2 in the previous example). This is the basis
of our method for efficiently selecting the most abundant
subisotopologues.

In order to generate the top $k$ isotopologues of a compound, we first
start by computing the top subisotopologues in decreasing order of
probability. A binary heap is utilized to retrieve the
subisotopologues in order; however, it is not populated with all
possible subisotopologues. Instead, when a subisotopologue is popped
from the heap, a new set of subisotopologues are put into the heap
based on the popped subisotopologues index tuple (a tuple which
describes how many of each isotope is in the subisotopologue).

First, we must find the most probable subisotopologue which amounts to
finding the mode of the multinomial, let $(x_1, x_2,\dots, x_m)$ be a
such a mode. Thus,
$P(x_1, x_2,\dots, x_m) \geq P(\dots,x_i+1,x_j-1,\dots)$ for any
indicies $x_i$ and $x_j$. Since $(x_1, x_2,\dots, x_m)$ is the mode and
$P(\dots,x_i+1,x_j-1,\dots) = P(x_1, x_2,\dots,
x_m)\cdot\frac{p_i\cdot( x_j-1)}{p_j\cdot( x_i+2)}$, it must be that
$\frac{p_i\cdot( x_j-1)}{p_j\cdot( x_i+2)} \leq 1$ for any indicies
$x_i$ and $x_j$. 

Furthermore, because
$P(\dots,x_i+(b+1),x_j-(b+1),\dots) =
P(\dots,x_i+b,x_j-b,\dots)\cdot\frac{p_j\cdot( x_k-b)}{p_k\cdot(
  x_j+(b+1))}$ and
$\frac{p_j\cdot( x_k-b)}{p_k( x_j+(b+1))} \leq \frac{p_j\cdot(
  x_k-(b-1))}{p_k\cdot( x_j+b)} \leq \dots \leq \frac{p_i\cdot(
  x_j-1)}{p_j\cdot( x_i+2)} \leq 1$ , we can see that
$P(\dots,x_i+b,x_j-b,\dots) \leq P(\dots,x_i+(b+1),x_j-(b+1),\dots)$
for any $b \leq \min(x_i,x_j)$. This means every time the
$i^\text{th}$ entry in the index tuple is increased and the
$j^\text{th}$ entry is decreased (thus moving further from the mode in
$\text{L}_1$ distance), the probability either decreases or stays the
same.

The relationship between $\text{L}_1$ distance from the mode and the
probability still holds when other index tuple entries have been
perturbed away from the mode.
$P(\dots,x_i+1,x_j-1,\dots) \geq P(\dots,x_i+2,x_j-1,x_k-1\dots)$
because
$P(\dots,x_i+2,x_j-1,x_k-1\dots) =
P(\dots,x_i+1,x_j-1,\dots)\cdot\frac{p_i\cdot( x_k)}{p_k( x_i+2)}$ and
$\frac{p_i\cdot( x_k)}{p_k( x_i+2)} \leq \frac{p_i\cdot(x_k-1)}{p_k(
  x_i+2)} \leq 1$.

Finally, we also have
$P(\dots,x_i+1,x_j-1,\dots) \geq
P(\dots,x_i+1,x_j-1,x_k+1,x_{\ell}-1\dots)$ because
$P(\dots,x_i+1,x_j-1,x_k+1,x_{\ell}-1\dots) =
P(\dots,x_i+1,x_j-1,\dots)\cdot\frac{p_k\cdot( x_\ell)}{p_\ell(
  x_k+1)}$ and
$\frac{p_k\cdot( x_\ell)}{p_\ell( x_k+1)} \leq \frac{p_k\cdot(
  x_\ell-1)}{p_\ell( x_k+2)} \leq 1$.

Since the probability never decreases as we move closer to the mode,
the mode can be found by moving in the direction of ascending
probability until a maximal position is reached. A strong position to
start can be found by finding the mode of the binomials which make-up
the multinomial. Note that this starting point may not be a valid
position if the entries in the index tuple do not sum to the total
size of the subisotopologue.

In order to populate the heap with the best possible next
subisotopologue, any subisotopologue in the heap must have all
subisotopologues between itself and the mode already in the heap (or
have been popped from the heap). This can be accomplished by pushing
all neighbors of the subisotopologue which has been popped from the
heap; however, there needs to be a system to make sure the same
subisotopologue is not proposed by multiple neighbors. This can be
done using a set, but a set requires memory and time spent checking to
see if an element is in the set. Instead, one only needs a proposal
scheme which can propose all subisotopologues in correct order without
any duplicates.

One such proposal method can be done by keeping track of the index of
the largest (in lexicographical order) entry in the index tuple which
has been incremented and the largest which has been decremented away
from the mode, let them be denoted $i'$ and $j'$, respectively. W.l.o.g.,
let $i' < j'$, then $x_i$ is the same as the mode for all
$i > i', i \neq j$. When an index tuple
$I=(x_1 + a_1, x_2 + a_2, \ldots x_{'i} + a_{i'},\ldots, x_{'j} +
a_{j'}, x_{'j+1},\ldots)$ (where $a_{i'}>0$ and $a_{j'}<0$ )is popped
from the heap, it proposes neighbors by taking all pairs of
$(i\geq i', j\geq j')$ and creating, then pushing into the heap, new
index tuples where the $i^\text{th}$ entry is incremented by one and
the $j^\text{th}$ entry is decremented by 1.

This proposal scheme means any index tuple may be proposed by only one
unique neighboring index tuple. For any index tuple $I$, the index
tuple which proposed it, $I'$ can be found by taking $i'$ and $j'$ of
$I$ and decrementing the entry at the ${i'}^\text{th}$ index by one
and incrementing the ${j'}^\text{th}$ entry by one. It can not be that
$I'$ incremented an index $i< i'$ to become $I$. If there was some
entry at index $i< i'$ incremented in $I'$ to become $I$, then the
${i'}^\text{th}$ entry in $I'$ would have been incremented previously;
however, incrementing $x_{i'}$ and then $x_i$ breaks the lexicographic
ordering so this is not allowed. The same logic holds for why only the
entry $x_{j'}$ in $I'$ can be decremented when proposing $I$. Since
each index tuple has only one unique index tuple which can propose it,
no index tuple may be proposed multiple times since the one who
proposes it can only be popped from the heap once.

For any valid index tuple $I$, the index tuple which proposed it can
be found by the method above. This is always done by decrementing an
entry whose index which is greater than the corresponding entry in the
mode and incrementing an entry which is less than the corresponding
entry in the mode, every time the previous index tuple is found we move
closer to the mode. This can only be done a finite amount of times
before the mode is reached. Therefore, from any valid index tuple
there is a unique path from the mode to the tuple using this proposal
scheme and so any valid index tuple is able to be proposed.

Once the top subisotopologues are generated, we want to perform
selection on the possible resulting isotopologues without first
calculating all isotopologues. We do so by implementing a modified
version of Serang's selection on $X+Y$.

\subsection{Selection on two subisotopologues}
Serang's method is modified in two major ways, the first is that the
pairwise selection is done on a list of mass and log-abundance
pairs. This list of pairs is not static. The second difference is that
this selection must be online so that we may always ask for the next
layer of isotopologues which have not been previously selected.

Serang's method works by putting layer products (a Cartesian product
of two layers of a LOH) into a heap either according to the minimum
value in the layer product, or, if the minimum value has been popped
already, according to the maximum. Once the minimum value is popped
all values in the layer product are calculated and inserted into a
list of candidate values for a final one-dimensional $k$-select. Layer
products are done popping when the total amount of candidate values
from layer products whose maximum has been popped is at least $k$.

In the online version, the layer products are done popping when the
total amount of candidate values from layer products whose maximum has
been popped is at least the amount of the cumulative selections
$k_1 + k_2 + \cdots = k$. The heap and the list of candidate values
are not modified between the multiple selections (except that the
smallest first $k_i$ values are removed from the list of candidate
values). Therefore, the layer products popped until the total amount
of candidate values from layer products whose maximum has been popped
is at least $k$ is the same regardless of whether it is one $k$-select
or a $k_1$- and a $k_2$-select.

In Serang's manuscript, in the proof of lemma 7 the author mentions
the possible size of the layer products added in phase 2 and says that
the layer products added in phase 2 for which either $u'=1$ or $v'=1$
is at most $2n \in O(n)$. In fact, the size of the layer is limited by
the fact that the previous layer along that axis must have had its
maximum value popped from the heap before the layer appended in phase
2 could have its minimum popped. Because of this, the added layer's
size is at most $\alpha\cdot s \in O(k)$ and so the lemma could
improve the statement by replacing $O(n + k)$ with $O(k)$.

\subsection{Selection on a compound}
\begin{figure}
  \centering
  \includegraphics[width=.5\textwidth]{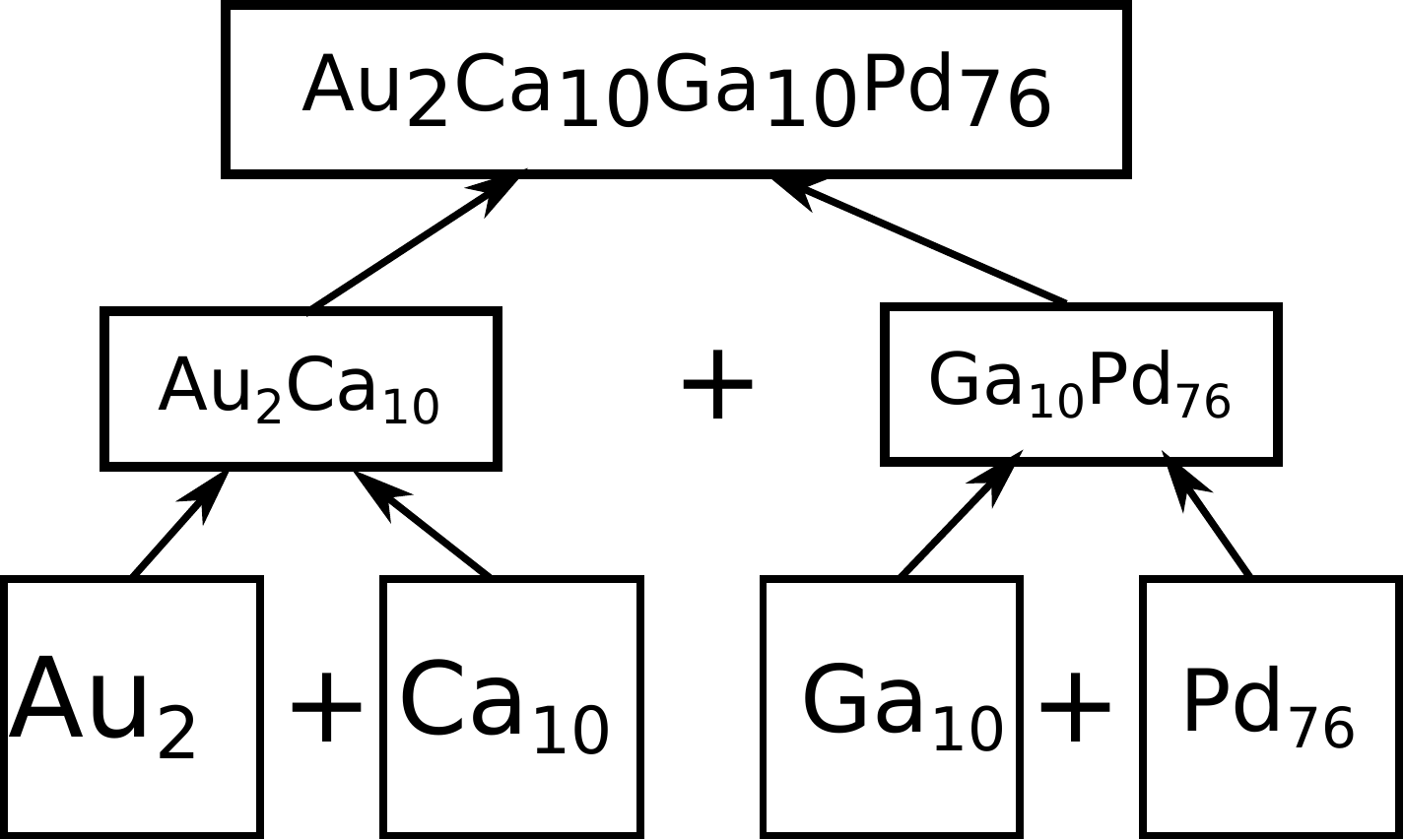}
  \caption{\textbf{Illustration of the balanced binary tree for
      Palladium alloy Pgc, \ch{Au2Ca10Ga10Pd76}.} The leaves are
    subisotopologue generators, one for each element. All nodes above
    the leaves combine their child compounds using the modified
    pair-wise selection from Serang's method. The root generates
    isotopologues of the human BRCA2 protein, and every other node
    generates isotopologues for some smaller constituent
    compound.\label{fig:tree}}
\end{figure}

The method described above is able to efficiently get the top $k$ many
combinations of two subisotopologues; however, for compounds of more
than two elements this method alone is unable to calculate any
isotopologues. In order to combine all subisotopologues, a balanced
binary tree of two different kinds of nodes is formed. The leaves of
the tree are all subisotopologue generators and all other nodes,
including the root, perform selection on $X + Y$
(Figure~\ref{fig:tree}).

At each layer of the tree above the leaves, pairs of smaller compounds
are merged and their top values are returned to the parent. To
retrieve the top $k$ isotopologues of the whole compound, the root is
asked to generate the top $k$ isotopologues from the combination of
its children. The root will initially have two empty lists, so it must
ask its children to produce values. Since we are using a method in
which the input lists are LOHified, a parent node will ask each child
to generate their next whole layer of values. Since every value in a
layer is at least as good as every value in the next layer, a child
only needs to generate the next layer when the output of the parent
has used all of the child's previously generated values.

Initially, every node in the tree will require both children to
generate values; however, the size of the first layer is always one so
this is not a lot of work. After the initial layer of each child is
generated, when the parent needs to generate more layers it only asks
one child at a time for more values. This is done by comparing the
maximum value in the last generated layers of each child and only
extending the child who has the smaller maximum value. As $k$
increases, when the root asks its children to generate more layers,
this request is unlikely to ripple down to the leaves because some
interior nodes will have previously generated layers in which not all
the values have been used.

\subsection{Time analysis}
If the leaves are removed from the tree, then the tree has the same
time complexity as FastSoftTree seen in \cite{kreitzberg:selection}.
The difference in the algorithms is that this algorithm uses LOHs
where FastSoftTree uses soft-heaps \cite{chazelle:soft}. LOHs and
soft-heaps have the same theoretical runtime for selection, but in
practice LOHs are significantly faster due to the data being
contiguous in memory. Therefore, the runtime of the tree formed from
$X+Y$ nodes is $\in O(m\cdot n + k\cdot m^{2\log_2(\alpha)})$ where $n$
is the number of subisotopologues generated and $m$ is the number of
elements in the compound. When generating subisotopologues, the leaf
nodes form tensors which have the same dimensionality as the number of
isotopes of the element. Thus, the subgenerators themselves have the
same time complexity as the SortTensor method used in Kreitzberg
\emph{et al.}, and therefore has time complexity
$\in O(n_e\cdot m_e + k_e\cdot m_e^2 + k_e\log(k_e \cdot n_e ))$, $n_e$
is the number of element $e$ in the compound, $m_e$ is the number of
isotopes of $e$, and $k_e$ is the number of subisotopologues
generated. For small $k$ the algorithm is leaf heavy, for large enough
$k$, most of the work done will be in the interior nodes and so the
tree becomes dominated by the $X+Y$ selections.

\section{Results}
Here, we compare our algorithm, which we call \textsc{NeutronStar},
versus \textsc{IsoSpec}. The \texttt{C++} interface for
\textsc{IsoSpec} was used with flags set so that only the masses and
log-probabilities are generated, specifically an instance of
TotalProbFixedEnvelope was created with flags (true, false, true,
true, false). Both \textsc{IsoSpec} and \textsc{NeutronStar} were
compiled with \texttt{g++ -O3 -march=native -mtune=native -std=c++17}.
The executables ran on a computer with dual AMD Epycs 7351 with 256GiB
of RAM.

\begin{figure}[H]
  \centering
  \includegraphics[width=.9\textwidth]{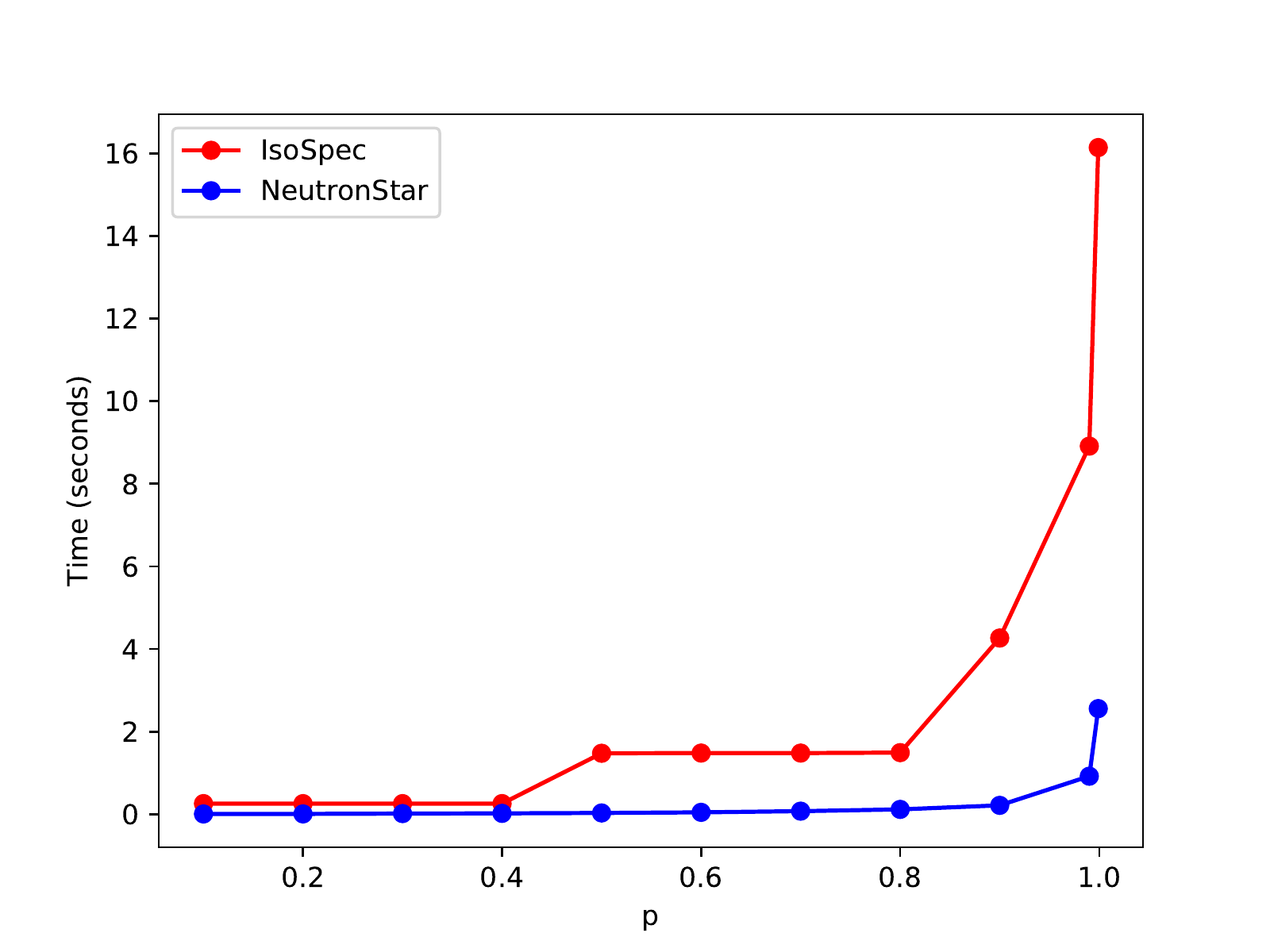}
  \caption{\textbf{Runtimes of \textsc{IsoSpec} and
      \textsc{NeutronStar} as $p$ (and, correspondingly, $k$ vary).}
    Each run was averaged over 10 iterations. Due to \textsc{IsoSpec}
    gathering isotopologues in batches and subsequently trimming
    excess, its runtimes may vary sharply as $p$ is changed. 50
    different $p$ values, evenly spaced from 0.01 to 0.95, were
    used.\label{fig:isospec-v-neutronstar}}
\end{figure}

\textsc{IsoSpec} generates a superset of the needed isotopologues and
then performs one-dimensional selection to retrieve the most
abundant. The isotopologues \textsc{IsoSpec} generates are chosen as a
function of $p$, the cumulative abundance threshold. To ensure a fair
comparison, we first run \textsc{IsoSpec} using a $p$, then use the
number of $k$ peaks it returns as the rank threshold for
\textsc{NeutronStar}. A runtime comparison between \textsc{IsoSpec}
and \textsc{NeutronStar} is performed in
Figure~\ref{fig:isospec-v-neutronstar}.

\subsection{Time}
\begin{table}[H]
\centering
  \small
\begin{tabular}{r|llll}
  Molecule & $p$ & $k$ & \textsc{IsoSpec}(s) & \textsc{NeutronStar}(s) \\
  \hline
  BRCA2 & 0.1 & 155717   & 0.069926  & 0.0168335 \\
           & 0.2 & 436377   & 0.0790409 & 0.0310062 \\
           & 0.3 & 868708   & 0.0809641 & 0.053721  \\
           & 0.4 & 1514835  & 0.0945277 & 0.0862687 \\
           & 0.5 & 2484628  & 0.0975662 & 0.135625  \\
           & 0.6 & 3984714  & 0.795611  & 0.212852  \\
           & 0.7 & 6465410  & 0.800833  & 0.344733  \\
           & 0.8 & 11135622 & 1.05385   & 0.577102  \\
           & 0.9 & 22935591 & 0.691333  & 1.11614   \\
           & 0.99 & 110918381 & 5.21918 & 5.38021 \\
           & 0.999 &  310931441 & 10.9851 & 14.2249 \\
  \hline
  Palladium & 0.1 & 9134    & 0.254897 & 0.0039866 \\
  alloy Pgc& 0.2 & 25806   & 0.254513 & 0.0050412 \\
           & 0.3 & 52855   & 0.254586 & 0.0128452 \\
           & 0.4 & 95387   & 0.255309 & 0.0185391 \\
           & 0.5 & 162857  & 1.47417  & 0.0280916 \\
           & 0.6 & 274344  & 1.47892  & 0.0432593 \\
           & 0.7 & 473917  & 1.47872  & 0.0720034 \\
           & 0.8 & 890269  & 1.48936  & 0.114372  \\
           & 0.9 & 2074266 & 4.26516  & 0.212619  \\
           & 0.99 & 13466926 & 8.91341 & 0.918616 \\
           & 0.999 & 47409787 & 16.1441 & 2.55547 \\
  \hline
  \ch{Xe50} & 0.1 & 2510   & 0.800637 & 0.0041821 \\
           & 0.2 & 6711   & 0.798412 & 0.0079587 \\
           & 0.3 & 12909  & 0.801227 & 0.0113697 \\
           & 0.4 & 21919  & 0.798453 & 0.0175081 \\
           & 0.5 & 35243  & 0.800154 & 0.0284915 \\
           & 0.6 & 55890  & 0.797688 & 0.0395307 \\
           & 0.7 & 91046  & 8.39407  & 0.0672331 \\
           & 0.8 & 159438 & 8.38149  & 0.115158  \\
           & 0.9 & 332449 & 8.38418  & 0.242873  \\
           & 0.99 & 1564230 & 35.8607& 1.15485 \\
  \hline
  \ch{Sn20Xe20Nd20Dy20} & 1e-12 & 1 & 0.000365 & 0.000158 \\
           & 1e-11 & 50 & segfaulted & 0.0002998 \\
           & 1.99685e-10  & 100  & ---- & 0.0003539 \\
           &  5.28451e-05 & 100000000  & ---- & 7.4452 \\
\end{tabular}
\caption{\textbf{Runtimes of \textsc{IsoSpec} and \textsc{NeutronStar}
    for three molecules, all with $\alpha=1.05$.} The first molecule
  is BRCA2 \ch{C16802H26738N4640O5411S121}, an organic molecule
  associated with a significant risk of breast cancer. The second
  molecule is palladium alloy Pgc, a dental amalgam with chemical
  formula \ch{Au2Ca10Ga10Pd76} and CID 6337993. The third compound is
  simply \ch{Xe50} to illustrate that the \textsc{NeutronStar}
  algorithm scales better than \textsc{IsoSpec} as the dimensionality
  of the subisotopologue increases. When \textsc{IsoSpec} segfaulted,
  larger $p$ were not run.\label{table:runtimes}}
\end{table}

\subsection{Space}
\begin{table}[H]
\centering
\small
\begin{tabular}{r|llll}
  Molecule & $p$ & $k$ & \textsc{IsoSpec} & \textsc{NeutronStar} \\
  \hline
  BRCA2 & 0.1 & 155717   & 96.1 MiB  & 6.0 MiB \\
           & 0.3 & 868708   & 96.1 MiB & 30.8 MiB \\
           & 0.5 & 2484628   & 96.1 MiB & 88.6 MiB  \\
           & 0.7 & 6465410  & 768.1 MiB & 232.1 MiB \\
           & 0.9 & 22935591  & 768.1 MiB  & 771.0 MiB  \\
  \hline
  Palladium & 0.1 & 9134    & 9.7 MiB      & 0.9751 MiB \\
  alloy Pgc& 0.3 & 52855    & 9.7 MiB      & 4.4 MiB \\
           & 0.5 & 162857   & 63.5 MiB     & 12.3 MiB \\
           & 0.7 & 473917   & 63.5 MiB     & 33.5 MiB \\
           & 0.9 & 2074266  & 419.7 MiB    & 115.0 MiB \\
           & 0.99 & 13466926  & 831.8 MiB  & 596.2 MiB \\
           & 0.999 & 47409787  & 3.1 GiB   & 1.8 GiB \\    
  \hline
  \ch{Xe50} & 0.1 & 2510    & 30.7 MiB & 1.4 MiB \\
           & 0.3 & 12909    & 30.7 MiB & 5.4 MiB \\
           & 0.5 & 35243    & 30.7 MiB & 21.6 MiB \\
           & 0.7 & 91046    & 166.0 MiB & 43.3 MiB  \\
           & 0.9 & 332449   & 166.0 MiB & 86.9 MiB  \\
           & 0.99 & 1564230 & 480.0 MiB & 352.3 MiB \\
\end{tabular}
\caption{\textbf{Memory usage of \textsc{IsoSpec} and
    \textsc{NeutronStar} (with $\alpha=1.05$) on three molecules.}
  Memory usage for the three compounds from
  Table~\ref{table:runtimes}.\label{table:memory}}
\end{table}

\subsection{Generated spectra}
\begin{figure}[H]
  \centering
  \includegraphics[width=1\textwidth]{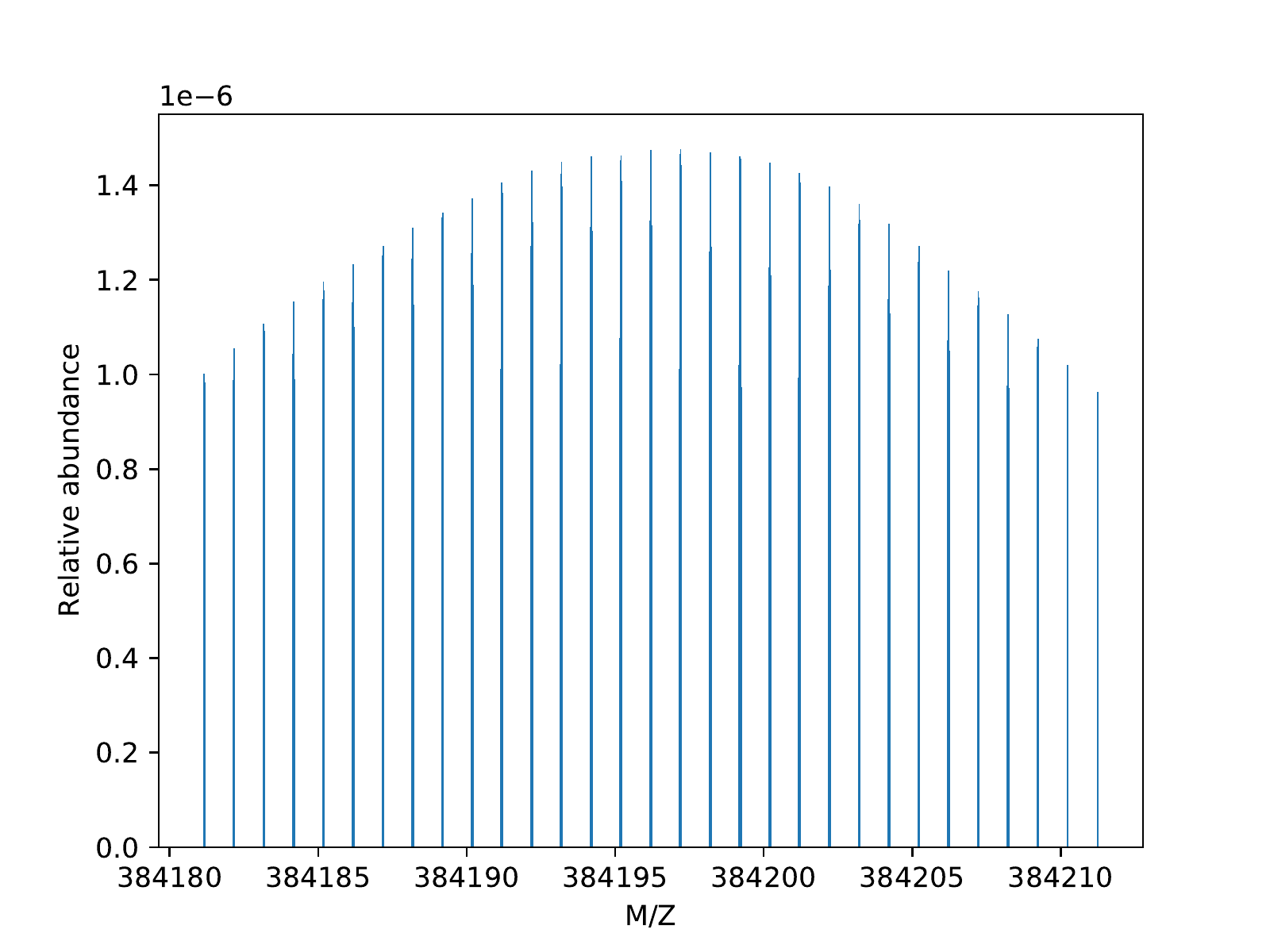}
  \caption{\textbf{Theoretical spectra of the top 10,000 peaks of
      BRCA2, \ch{C16802H26738N4640O5411S121}.} The top 10,000
    isotopologue peaks were generated by \textsc{NeutronStar} using
    $\alpha=1.05$. \textsc{NeutronStar} took 0.004748 seconds to
    generate the peaks and they cover a cumulative abundance of
    0.0109297.\label{fig:generated-spectra}}
\end{figure}

Figure~\ref{fig:generated-spectra} depicts the most abundant 10,000
peaks of BRCA2, \ch{C16802H26738N4640O5411S121}\cite{expasy:brca2}; at
a high resolution, these peaks are subtley staggered from one
another. This would be seen by a high mass accuracy spectrometer.

\subsection{Influence of $\alpha$ on runtime}
\begin{table}[H]
\centering
\small
\begin{tabular}{r|lll}
  Molecule & k & $\alpha$ & Time(s) \\
  \hline
  \ch{C16802H26738N4640O5411S121}           & 1000000 & 1.00  & 0.516298  \\
           & 1000000 & 1.05  & 0.0589054 \\
           & 1000000 & 1.10  & 0.0606883 \\
           & 1000000 & 1.15  & 0.067031  \\
           & 1000000 & 1.20  & 0.0635678 \\
           & 1000000 & 1.25  & 0.0697309 \\
           & 1000000 & 1.30  & 0.0871132 \\
           & 1000000 & 1.35  & 0.0747897 \\
           & 1000000 & 1.40  & 0.0884976 \\
           & 1000000 & 1.45  & 0.111566  \\
           & 1000000 & 1.50  & 0.103195  \\
           & 1000000 & 1.55  & 0.118836  \\
           & 1000000 & 1.60  & 0.105935  \\
           & 1000000 & 1.65  & 0.101384  \\
           & 1000000 & 1.70  & 0.136349  \\
           & 1000000 & 1.75  & 0.155706  \\
           & 1000000 & 1.80  & 0.148027  \\
           & 1000000 & 1.85  & 0.154379  \\
           & 1000000 & 1.90  & 0.170928  \\
           & 1000000 & 1.95  & 0.16922   \\
           & 1000000 & 2.00  & 0.128749  \\
\end{tabular}
\caption{\textbf{Relationship between $\alpha$ and the runtime for
    $\alpha \in [1,2]$.} The time reported is the average over 10
  iterations and all times reported are only from
  \textsc{NeutronStar}. If $\alpha=1$ then sizes of the layers do not
  increase and so a layer-ordering with $\alpha=1$ is the same as
  sorting. The runtime is at its worst when the $\alpha=1$ and best
  when $\alpha=1.05$.\label{table:alpha-vs-runtime}}
\end{table}

\section{Discussion}
As seen in Table~\ref{table:runtimes}, for organic molecules of
moderate size (and therefore similar molecules whose elements have a
smaller amount of isotopes), \textsc{NeutronStar} is is roughly
equivalent to \textsc{IsoSpec}, the faster algorithm is largely
determined by the $p$ parameter. On the toy molecule
\ch{Sn20Xe20Nd20Dy20}, \textsc{IsoSpec} was unable to run on $p=1e-11$
because it caused a segfault (on a machine with 256GB of RAM) whereas
\textsc{NeutronStar} was able to perform the selection in 0.0002998
seconds. The ability of \textsc{NeutronStar} to handle
subisotopologues with a large amount of isotopes will become
increasingly more important for large molecules because, given enough
copies of one element, even trace isotopes may appear. For
\textsc{NeutronStar}, the choice of $\alpha$ can also significantly
impact the runtime as seen in Table~\ref{table:alpha-vs-runtime}.

The significant advantage \textsc{NeutronStar} has for large molecules
is possibly due to the method \textsc{IsoSpec} uses to propose the
neighbors of the popped isotopologues. Since it proposes a neighbor
for every subisotopologue, the space of the index tuples proposed will
form a tensor with dimension equal to the number of subisotopologues.
In \textsc{NeutronStar} the space of isotopologues proposed can only
form a matrix; however, there will be multiple matrices but the
overall the memory usage for \textsc{NeutronStar} should be lower than
\textsc{IsoSPec}. Also, \textsc{IsoSPec} requires a set of index
tuples in order to avoid proposing duplicate tuples. For large
problems this can cause significant overhead. The set is avoided in
\textsc{NeutronStar} due to the proposal scheme.

When comparing the peaks of \textsc{NeutronStar} and \textsc{IsoSpec},
they tend to agree to 15 significant figures (often reporting perfect
matches) in their mass and 10 significant figures in their relative
log-abundance, even for peaks whose masses match perfectly. The
difference on the level of agreement between the two quantities may be
a result of the use of the Stirling approximation by \textsc{IsoSpec},
but the overall disagreement is likely due to \textsc{IsoSpec}
calculating the values for each isotopologue independently instead of
carrying the values along with the subisotopologues as done in
\textsc{NeutronStar}.

Currently, \textsc{NeutronStar} is not configured to report the
isotopic make-up of the resulting isotopologues. This could be
modified by keeping track of the index tuple as an isotopologue is
created while climbing up the binary tree; however this would result
in a considerable performance reduction and remove one of the more
novel aspects of this algorithm. Furthermore, since each $X+Y$
selection node in the binary tree can do online selection, the
interface in \textsc{NeutronStar} can be adjusted to allow the user to
do multiple selections without having to recompute any previously
reported isotopologues. It is also easy to modify the \texttt{C++}
code to accept the same parameter $p$ as \textsc{IsoSpec} does. The
authors believe $k$ is the desired parameter because the amount of
peaks generated given $p$ is hard to estimate which, on one extreme,
can lead to segfaults and on the other to repeatedly retrieving a very
small set of peaks.

The algorithm can easily be adapted to doing selection on $X_1 + X_2 +
\cdots + X_m$ by removing the subisotopologue generators and replacing
them with LOH generators which take a list of values as a
parameter. An efficient selection on $X_1 + X_2 + \cdots + X_m$ may be
useful for solving a certain families of ILPs or set-cover problems
and their applications (\emph{e.g.}, protein inference).

In the future, it may be possible to use LOHs inside the
subisotopologue generators similar to the $X+Y$ selection nodes. This
could be a considerable speed-up because it avoids the
$\Omega(n\log(n))$ bounds created by sorting the subisotopologues. The
downside is that not every index tuple in a layer product will be
valid. Since the subisotopologue generators perform a different task
than the $X+Y$ nodes, it may be beneficial for each node type to have
different $\alpha$ values.

\section{Acknowledgements}
This work was supported by grant number 1845465 from the National
Science Foundation.

\section{Supplemental information}
The \textsc{NeutronStar} algorithm, implemented in \texttt{C++}, can
be found freely at
\url{https://bitbucket.org/orserang/neutronstar/}.

\end{document}